# When Can we Call a System Self-organizing?

Carlos Gershenson and Francis Heylighen
Centrum Leo Apostel, Vrije Universiteit Brussel
Krijgskundestraat 33, Brussels, 1160, Belgium
{cgershen, fheyligh}@vub.ac.be
http://www.vub.ac.be/CLEA

**Abstract**. We do not attempt to provide yet another definition of self-organization, but explore the conditions under which we can model a system as self-organizing. These involve the dynamics of entropy, and the purpose, aspects, and description level chosen by an observer. We show how, changing the level or "graining" of description, the same system can appear self-organizing or self-disorganizing. We discuss ontological issues we face when studying self-organizing systems, and analyse when designing and controlling artificial self-organizing systems is useful. We conclude that self-organization is a way of observing systems, not an absolute class of systems.

## 1 Introduction

There have been many notions and definitions of self-organization, useful in different contexts (for a non-technical overview, see [1]). They have come from cybernetics [2, 3, 4, 5], thermodynamics [6], mathematics [7], information theory [8], synergetics [9], and others. Many people use the term "self-organizing", but it has no generally accepted meaning, as the abundance of definitions suggests. Also, proposing such a definition faces the philosophical problem of defining "self", the cybernetic problem of defining "system", and the universal problem of defining "organization". We will not attempt to propose yet another definition of self-organizing systems. Nevertheless, in order to try to understand these systems better, we will explore the following question: which are the necessary conditions in order to call a system "self-organizing"? We do so by combining insights from different contexts where self-organizing systems have been studied.

The understanding of self-organizing systems is essential for the artificial life community, because life obviously exhibits self-organization, and so should the systems simulating it. As a most basic example, we can look at swarming or herding behaviour, in which a disordered array of scattered agents gathers together into a tight formation. Intuitively, it seems that the system of agents has become more "organized". But precisely what does this mean?

In the following section we explore the role of dynamics in self-organizing systems. We provide examples of systems that are self-organizing at one level but not at another one. In Section 3 we note the relevance of the observer for perceiving self-organization. We discuss some deeper conceptual problems for understanding self-organizing systems in Section 4. In Section 5 we discuss applications in artificial self-organizing systems, and when using this approach is appropriate. We draw concluding remarks in Section 6.

## 2 The representation-dependent dynamics of Entropy

A property frequently used to characterize self-organization is an increase of order which is not imposed by an external agent (not excluding environmental interactions) [1]. The most common way to formalize the intuitive notion of "order" is to identify it with the negative of *entropy*. The second law of thermodynamics states that in an isolated system, entropy can only decrease, not increase. Such systems evolve to their state of maximum entropy, or thermodynamic equilibrium. Therefore, self-organizing systems cannot be isolated: they require a constant input of matter or energy with low entropy, getting rid of the internally generated entropy through the output of heat ("dissipation"). This allows them to produce "dissipative structures" which maintain far from thermodynamic equilibrium [6]. Life is a clear example of order far from thermodynamic equilibrium.

However, the thermodynamical concept of entropy as the dissipation of heat is not very useful if we want to understand information-based systems, such as those created by ALife modellers. For that, we need the more general concept of *statistical entropy* (H) which is applicable to any system for which a state space can be defined. It expresses the degree of uncertainty we have about the state *s* of the system, in terms of the probability distribution *P(s)*.

$$H(P) = -\sum_{s \in S} P(s) \log P(s) \qquad (1)$$

In this formulation, the second law of thermodynamics can be expressed as "every system tends to its most probable state" [4]. This is in as sense a tautological law of nature, since the probabilities of the states are determined by us according to the tendencies of systems. At a molecular level, the most probable state of an isolated system is that of maximum entropy or thermodynamic equilibrium, where the molecules are distributed homogeneously, erasing any structure or differentiation. But does this apply as well to a real or artificial living organism?

We have to be aware that probabilities are relative to a level of observation, and that what is most probable at one level is not necessarily so at another. Moreover, a state is defined by an observer, being the conjunction of the values for all the variables or attributes that the observer considers relevant for the phenomenon being modelled. Therefore, we can have different degrees of order or "entropies" for different models or levels of observation of the same entity.

Let us illustrate this with the following, very simple example. Consider a system with four possible "microstates", *a1*, *a2*, *b1*, and *b2*, at the lowest, most detailed level of description. At the higher, more abstract level of description, we aggregate the microstates two by two, defining two macrostates: $A = \{a1, a2\}$ and $B = \{b1, b2\}$. This means that the system is in macrostate *A* if it is either in microstate *a1* or in microstate *a2*. The probabilities of the macrostates are simply the sum of the probabilities of their sub-states. Let us suppose that we start from an initial probability distribution so that $P(a1) = P(b1) = 0.1$ and $P(a2) = P(b2) = 0.4$. This implies $P(A) = P(B) = 0.5$. We can calculate the statistical entropy *H* using (1), getting $H \approx 1.72$ at the lower level, and $H=1$ at the higher level.

Now consider a second distribution $P(a1) = P(a2) = 0.2$ while $P(b1) = P(b2) = 0.3$. Therefore, $P(A) = 0.4$ and $P(B) = 0.6$. Now we have $H \approx 1.97$ at the lower and $H \approx 0.97$ at the higher levels. Subtracting the initial *H* from the second, we have $\partial H/\partial t \approx 0.24$ at a lower level and $\partial H/\partial t \approx -0.029$ at the higher level. We have a change of distribution where entropy is *decreased* at the higher level ("self-organization"), and *increased* at the lower level ("self-disorganization"). To get the inverse change, we can just assume the final states to be the initial ones and vice versa. We would then have self-organization at the lower level and self-disorganization at the higher level.

This can be represented graphically in Figure 1, where tones of gray represent the probabilities of the states (darker colour = lower value). The macrostates provide a coarse-grained [10] representation of the system, while the microstates provide a fine-grained one. We can visualize entropy as homogeneity of colours/probabilities. At the lower level, the distribution becomes more homogeneous, and entropy increases. At the higher level, the distribution becomes more differentiated.

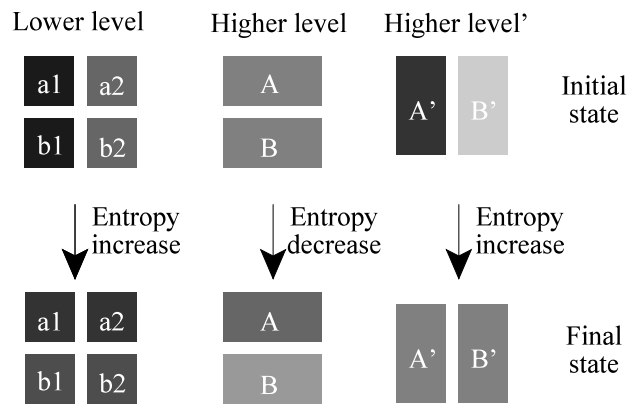

Figure 1. Entropy, seen as homogeneity, increases at lower level, while it might increase or decrease at the higher level, depending on how we divide the states

Entropy not only depends on higher or lower levels of abstraction, but also on how we set the boundaries between states. Let us define alternative macrostates *A'*, which has *a1* and *b1* as sub-states, and *B'*, with *a2* and *b2* as sub-states. Using the same values for the probabilities, we see that for the alternative macrostates the initial $H\approx0.72$ and the final $H=1$. So we have that $\partial H/\partial t \approx 0.27$, which means that the statistical entropy increases in this macrorepresentation, while it decreases in the previous one. The system appears to be self-organizing or disorganizing, depending not only on the level at which we observe it, but also on *how* we do the "coarse-graining" of the system, that is to say, which variables we select to define the states.

The variables defined by the values (*A*, *B*), respectively (*A'*, *B'*), represent two aspects of the same system, where the observer has focussed on different, independent properties. For example, a particle's state includes both its position in space and its momentum or velocity. A subsystem is defined as a physical part of a system, limited to some of its components. Similarly, an *aspect system* can be defined as a *functional part of a system*, limited to some of its properties or aspects [11].

Let us illustrate this with a typical ALife model: swarming behaviour. Groups of agents can be seen as subsystems of the swarm. The positions of all agents define one aspect system, while their velocities define another aspect system. Assume we start with non-moving agents scattered all over the simulated space. The position aspect is characterized by maximum entropy (agents can be anywhere in space), while the velocity aspect has minimum entropy (all have the same zero velocity). According to typical swarming rules, the agents will start to move with varying speeds towards the centre of the swarm while mutually adjusting their velocities so as not to bump into each other. This means that their states become more concentrated in position space, but more diffuse in velocity space. In other words, entropy decreases for the positions, while increasing for the velocities. Depending on the aspect we consider, the swarm self-organizes or self-disorganizes!

This example may appear too specific to support our argument. Let us therefore show that dynamical processes in general exhibit this kind of aspect-dependent or level-dependent behaviour, either increasing or decreasing entropy. A dynamical system, where every state is mapped deterministically onto a single next state, can be seen as special case of a Markov process, where a state $s_i$ is mapped stochastically onto any number of other states $s_j$ with a fixed transition probability $P(s_i \rightarrow s_j)$. To turn a deterministic dynamics into a stochastic one, it suffices to apply coarse-graining, aggregating a number of microstates into a single macrostate. Transitions from this macrostate can now go to any other macrostate that includes one of the microstates that were the initial destinations of its microstates.

It can be proven that the statistical entropy of a distribution cannot decrease if and only if the Markov process that describes the mapping from initial to final distribution is *doubly stochastic* [12]. This means that for the matrix of transition probabilities between states, the sum over a row and the sum over a column must be one. The sum over a row (probability of a given state ending in any state of the state space) is necessarily one, by definition of probability. However, the sum over a column is not a probability but can be seen as a measure of the "attractiveness" or "fitness" F of a state s:

$$F(s_i) = \sum_j P(s_j \rightarrow s_i) \qquad (2)$$

High fitness ($F > 1$) of a state $s$ means that on average more transitions enter $s$ than leave $s$ (the sum of all transition probabilities leaving $s$ can never be >1). Thus, while the process runs, high fitness states become more probable, and low fitness states less probable, increasing their differentiation, as would be expected from a process undergoing self-organization.

A doubly stochastic process is defined by the requirement that $F(s_i) = 1$ for all states $s_i$. This corresponds to the entropy increasing processes studied in traditional thermodynamics. In a more general process, probability eventually concentrates in the high fitness states, decreasing overall entropy if the initial distribution is more homogeneous, increasing it if it is more "peaked". Therefore, a necessary and sufficient condition for a Markov process to allow self-organization is that it has a non-trivial fitness function, *i.e.* there exist states such that $F(s) \neq 1$. The sum of all probabilities must be 1. Therefore, for every $F(s)>1$, there must be at least one $F(s')<1$. This condition is equivalent to saying that the dynamics has a differential "preference" for certain states $s$ over other states $s'$. Any dynamics that allows attractors has such an inbuilt preference for attractor states over basin states [1]. This is the most general case, and the one typically found in ALife simulations.

For any initial and final distribution of probabilities, such as the ones in the examples we discussed, it is possible to determine a matrix of transition probabilities that maps the one onto the other. This matrix will in general not be doubly stochastic, and therefore allow self-organization as well as disorganization. Therefore, the same system, described in different aspects or levels of abstraction can be modelled as self-organizing in the one case, as self-disorganizing in another.

## 3 The Role of the Observer

We have to be aware that even in mathematical and physical models of self-organizing systems, it is the *observer* who ascribes properties, aspects, states, and probabilities; and therefore entropy or order to the system. But organization is more than low entropy: it is structure that has a function or *purpose* [13]. Stafford Beer [4] noted a very important issue: what under some circumstances can be seen as organization, under others can be seen as disorder, depending on the purpose of the system. He illustrates this idea with the following example: When ice cream is taken from a freezer, and put at room temperature, we can say that the ice cream disorganizes, since it loses its purpose of having an icy consistency. But from a physical point of view, it becomes more ordered by achieving equilibrium with the room, as it had done with the freezer[1]. Again, the purpose of the system is not an objective property of the system, but something set by an *observer*.

W. Ross Ashby noted decades ago the importance of the role of the observer in relation to self-organizing systems: "A substantial part of the theory of organization will be concerned with *properties that are not intrinsic to the thing but are relational between observer and thing*" ([3], p. 258, emphasis in original).

Of course there should be a correlate in the world to the observations. The question now is: How frequent is this correlate, so that we can observe self-organization? What we need is a collection of elements that interact. By generalizing the second law of thermodynamics, we can see that the system through time will reach a more "probable" or "stable" configuration. We can say that it will reach an equilibrium or attractor[2]. The observer then needs to focus his/her viewpoint, in order to set the *purpose* of the system so that we can see the attractor as an "organized" state and to see it at the right *level* and in the right *aspect*, and then self-organization will be observed. We can see that this is much more common than what intuition tells us. Not only lasers, magnets, Bénard rolls, ant colonies, or economies can be said to be self-organizing. Even an ideal gas can be said to be self-organizing, if we say (contrary to thermodynamics) that the equilibrium state where the gas homogeneously fills its container, is "ordered" [4]. *Any* dynamical system *can* be said to be self-organizing [3]. *Self-organization is a way of modelling systems, not a class of systems*. This does not mean that there is no self-organization independently of the observer, but rather that self-organization is everywhere.

Of course, not all systems are usefully described as self-organizing. Most natural systems can be easily fit into the class "self-organizing", unlike the simple mechanisms we find in physics textbooks. Most artificial systems are hard to see as self-organizing. Many are not dynamic, others involve only one element (actually no system), and most of the rest follow sequences of rules that can be easily understood. Therefore there is no need to explain their functioning with the rather esoteric concept of "self-organization".

We have said that any dynamical system, if observed "properly", can be seen as self-organizing. But if we set a different purpose or description level, then *any* dynamical system can be self-disorganizing. An economy will not be seen as self-organizing if we look only at a short timescale, or if we look at the scale of only one small business. An ant colony will not be self-organizing if we describe only the *global* behaviour of the colony (*e.g.* as an *element* of an ecosystem), or if we only list the behaviours of individual ants. We have to remember that the description of self-organization is partially, but strongly, dependent on the observer.

---
[1] Thermodynamic entropy can be seen as order or disorder in different situations (*e.g.* [6, 4]. This also occurs with information entropy, as the debate between Wiener and Shannon showed.

[2] In some chaotic systems, this can take practically infinite time. But as systems approach an attractor, we can say that they follow this law.

## 4 Ontological Issues

One of the most common problems when discussing self-organizing systems is the meaning of emergence. Self-organizing systems typically have higher level properties that cannot be observed at the level of the elements, and that can be seen as a product of their interactions (more than the sum of the parts). Some people call these properties *emergent*. The problem we face is ontological. According to Aristotelean logic, a system cannot be more than one thing at a time. In this case, a system cannot be at the same time a set of elements and a whole with emergent properties. But by introducing an ontological distinction [14], we can clarify the issue.

We can distinguish two types of *being*: relative and absolute. The relative (*rel*-being) is experienced by an observer with a finite cognitive capacity. It therefore depends on her/his context, and is limited. Strictly speaking, every cognizer has a different *rel*-being of anything, since every cognizer has a different context. Theoretically, we can assume that there exists an absolute being (*abs*-being), which would be "the real thing" (Kant's *Ding-an-sich*), independent of the observer, which observers correlate to their *rel*-beings. We can observe any *abs*-being from an infinity of perspectives and describe an infinity of potential properties or aspects. Nevertheless, most *rel*-beings and contexts are similar, since they are inspired by the same *abs*-being seen by similar observers from a similar point of view. This enables us to share knowledge, but it is because of the different nuances in the different *rel*-beings and contexts that we fail to agree in every situation.

We can then say that the observation of a system at different abstraction levels or in different aspects is merely a difference in the perspective, and therefore the system *rel*-is different (only for the observers). But the system *abs*-is the same thing itself, independently of how we describe it. We can observe a cell as *rel*-being a bunch of molecules or as *rel*-being a living structure. But it *abs*-is both and even more. *Rel*-beings can be seen as different models or metaphors for describing the same thing. A change in the metaphor does not change the thing. If we define emergence as a process that requires a change of the model [15] in order to better understand and predict the system [8], then it becomes clear that there is no magic. Any dynamical system *abs*-is self-organizing and self-disorganizing at the same time, in the sense that it can potentially *rel*-be both.

Another confusion may arise when people describe systems as the lower level *causing* change in the emergent properties. Vice-versa, downward causation is the idea that higher level properties constrain or control components at the lower level [16]. Speaking about causality between abstraction levels is not accurate [14], because actually they *abs*-are the same thing. What we could say is that when we observe certain conditions in the lower level, we can expect to observe certain properties at a higher level, and vice versa. There is correlation, but not actual causation.

This leads us to what is probably the most fundamental problem. If we can describe a system using different levels, aspects, or representations, which is the one we should choose? As Prem [17] suggests, the level should be the one where the prediction of the behaviour of the system is easiest; in other words, where we need least information to make predictions[3] [8]. A possible way to formalize this requirement is by choosing the representation that minimizes the conditional entropy, *i.e.* the average uncertainty of the next state given the present state [5]. We can speculate that this is possible because of regularities in the system at that particular level, and that this is what leads people to try to describe how the simple properties cause the not so simple ones, either upward or downward.

## 5 Artificial Self-organizing Systems

Independently of the definition of self-organizing systems, if we see them as a perspective for studying systems, we can use this perspective for designing, building, and controlling systems. A key characteristic of an artificial self-organizing system is that structure and function of the system "emerge" from interactions between the elements. The purpose should not be explicitly designed, programmed, or controlled. The components should *interact* freely with each other and with the environment, mutually adapting so as to reach an intrinsically "preferable" or "fit" configuration (attractor), thus defining the purpose of the system in an "emergent" way [13].

Certainly this is not the only approach for designing and controlling systems, and in many cases it is not appropriate. But it can be very useful in complex systems where the observer cannot *a priori* conceive of all possible configurations, purposes, or problems that the system may be confronted with. Examples of these are organizations (corporations, governments, communities), traffic control, proteomics, distributed robotics, allocation of ecologic resources, self-assembling nanotubes, and complex software systems [13], such as the semantic web.

For artificial life, we believe that the perspective of self-organizing systems is essential, because we cannot explain the emergence, evolution, and development of life, whether *in vivo* or *in silico*, by restricting our models to a single level of abstraction. We need to understand how properties such as life, autocatalysis, or autopoiesis can emerge from interactions of elements without these properties, or how species or social properties can arise from individual interactions. This can only be done from a perspective of self-organizing systems. Therefore, it is important that the issues we have discussed are taken into account by artificial life researchers.

## 6 Conclusions

We proposed that self-organizing systems, rather than a *type* of systems, are a *perspective* for studying, understanding, designing, controlling, and building systems. This perspective has advantages and disadvantages, and there are systems that benefit from this approach, and others for which it is redundant. But even in the general case when the systems dynamics allows self-organization in the sense of entropy decrease, the crucial factor is the *observer*, who has to describe the process at an appropriate *level(s)* and *aspects*, and to define the *purpose* of the system. All these "make" the system to be self-organizing. In that sense, self-organization can be everywhere: it just needs to be observed.

---

[3]This argument could be also followed to decide which "graining" to choose.

We believe that this discussion on when and how to best model a system as self-organizing should be carried further in the artificial life community, since we all study and build systems from a self-organizing perspective. This would benefit not only the community, but every domain where the notion of self-organization is useful.